\documentclass[%
 reprint,
 amsmath,amssymb,
 aps,
]{revtex4-2}

\usepackage{graphicx}% Include figure files
\usepackage{dcolumn}% Align table columns on decimal point
\usepackage{bm}% bold math
\usepackage{verbatim}
\usepackage{cancel}
\usepackage{amsmath}
\usepackage{color}
\usepackage{tikz}
\usetikzlibrary{tikzmark}

\begin{document}

\preprint{APS/123-QED}

\title{Orthonormal Bases for Reconstructing Pairwise Interpulsar \\Correlations in Pulsar Timing Arrays}

\author{Dustin R. Madison}
 \email{dustin.madison@nanograv.org}
\affiliation{Department of Physics \& Astronomy, University of the Pacific, 3601 Pacific Avenue, Stockton, CA 95211, USA}

%\author{Colleagues}
%\affiliation{NANOGrav}

\date{\today}

\begin{abstract}
For pulsar timing arrays (PTAs), the telltale signature of an isotropic stochastic background of gravitational waves is a pattern of pairwise interpulsar timing correlations approximately following the Hellings \& Downs (HD) curve. Certain systematic errors and new physics processes also lead to interpulsar correlations with different patterns that can be distinguished from the HD curve to varied degrees. As evidence of HD correlations in PTA data mounts in coming years, it is important to develop principled strategies for flexibly and optimally reconstructing the pattern of interpulsar timing correlations, both to test how well the correlations track the HD pattern and to possibly detect additional effects, systematic or otherwise. To this end, we develop orthonormal basis functions that fully capture HD correlations and eliminate covariances between the HD curve and any additional correlated structure. We do this analytically and in a data-adaptive way informed by ``optimal statistic" analysis techniques widely used by PTA groups. These bases are adaptive in that they will vary from PTA to PTA and from data release to data release as new pulsars of varied timing quality and baseline are added to arrays, as instrumentation advances, and as observations accrue. We discuss how the techniques we introduce can be extended to future multi-signal searches by PTAs and to robust assessment of HD detection significance. 
\end{abstract}

\maketitle

\section{\label{sec:Intro}Introduction}
Pulsar timing arrays (PTAs) are decade-spanning efforts to precisely time dozens of pulsars and to detect in their timing behavior subtle correlated patterns with exciting causes, chief among them being ultra-low frequency gravitational waves (GWs). These ambitious projects have been diligently pursued by groups of (mainly radio) astronomers all over the globe \cite{rbc+19,dcl+16,h13,l16,jab+18,msb+23,vlh+16} demonstrating ever-improving scientific capabilities. 

A PTA with $N_p$ pulsars has $N=N_p(N_p-1)/2$ pairs of pulsars. The $i$th of these $N$ pulsar pairs has an angular separation $\theta_i$ from the perspective of the observer. GWs in the vicinity of the solar system induce correlated timing variations in each pair of pulsars \cite{ew75}. For a universe containing a large number of isotropically distributed GW sources, the amount of cross-correlated timing residual power in the $i$th pair of pulsars will depend on $\theta_i$ and be proportional to the value of the Hellings \& Downs (HD) curve \cite{hd83}:
\begin{eqnarray}
\Gamma(\theta_i)&=&\frac{1}{2}-\frac{(1-\mu_i)}{8}+\frac{3(1-\mu_i)}{4}\ln{\left(\frac{1-\mu_i}{2}\right)},
\end{eqnarray}
where $\mu_i\equiv\cos{\theta_i}$. The scale of the correlation will depend on the amplitude and spectral characteristics of the signal, but the variation with $\theta_i$ will follow $\Gamma$. The angular pattern of the correlations will change for anisotropic source populations and as the size of the source population shrinks \cite{cs13,ar22}. 

Multiple PTAs recently reported compelling observational evidence for interpulsar correlations consistent with the HD curve \cite{rzs+23,xcg+23,aaa+23_d,aaa+23_a}. The sources generating the GW background that is likely inducing these HD correlations is still an open question, but the amplitude and spectrum of the background are broadly consistent with what is expected from an ensemble of supermassive black hole binaries \cite{aaa+23_b,aaa+23_c}.

The HD curve is a type of overlap reduction function (ORF) that describes the anticipated fractional correlation between pairs of detectors in the presence of a signal of interest. Specifically, the HD curve is the ORF generated by a confused isotropic background of general relativistic GWs. But different physical phenomena can generate correlations consistent with different ORFs. Two alternative ORFs commonly considered by PTAs are monopoles (no dependence on $\theta_i$) and dipoles (proportional to $\cos{\theta_i}$) possibly arising from inaccurate timekeeping and errors in referencing pulse arrival times to the quasi-inertial solar system barycenter, respectively \cite{thk+16}. Additionally, many alternative theories of gravity admit non-Einsteinian GW polarization modes which give rise to ORF structure that differs from the HD curve \cite{cs12,grt15}. The space of interpulsar correlations for a PTA is an important forum for error checking and scientific exploration. As such, flexible analysis tools need to be developed and put to use. We aim to develop some such tools in this work.

In Section II, we discuss a common way in which pairwise interpulsar correlations are modeled as linear combinations of Legendre polynomials. In Section III, we construct an alternative basis for modeling interpulsar correlations that is specifically tailored to PTA searches for HD signal. In Section IV, we introduce an adjusted method for constructing this HD-oriented basis that is adaptive to the realities of PTA data quality and analysis methods. In Section V, we discuss how our methods can be extended to searches for multiple signals at once and walk through a specific example. We offer some concluding thoughts in Section VI.

%%%%%%%%%%%%%%%%%%%
%%%%%%%%%%%%%%%%%%%
%%%%%%%%%%%%%%%%%%%
\section{Decompositions of Interpulsar Correlations}
The HD curve can be written as a linear combination of Legendre polynomials as follows \cite{grt+14}:
\begin{eqnarray}
\Gamma(\theta)=\sum_{l=0}^\infty g_lP_l(\mu),
\end{eqnarray}
where $g_0=g_1=0$ and for $l\geq2$,
\begin{eqnarray}
g_l=\frac{3}{2}(2l+1)\frac{(l-2)!}{(l+2)!}.
\end{eqnarray}
These coefficients monotonically decrease and for large values of $l$, $g_l\sim l^{-3}$. The largest value is $g_2=5/16$ and, consequently, the HD curve is commonly described as quadrupolar. It nonetheless has power in all higher multipoles. 

The motivation for this sort of decomposition of $\Gamma$ comes from some linear algebra considerations. Legendre polynomials can be thought of as vectors in the infinite-dimensional vector space of square-integrable functions on the interval $I=[-1,1]$, often called $L^2(I)$. Furthermore, they are orthogonal with respect to the following inner product:
\begin{eqnarray}
\langle P_k,P_l\rangle_C&=&\int_{-1}^1P_k(\mu)P_l(\mu)d\mu,\nonumber\\
&=&\frac{2}{2l+1}\delta_{kl}.
\end{eqnarray}
Note that $\delta_{kl}$ is the Kronecker delta and that the ``$C$" in the above notation indicates that this is an inner product of ``continuous" functions; we will employ a discrete inner product, $\langle~,~\rangle_D$, in later sections. This integral property of the Legendre polynomials allows one to express any element of $L^2(I)$ as a linear combination of the $P_l$ elements. The full set of $P_l$ elements thus forms an orthogonal basis for $L^2(I)$, and since $\Gamma$ can be expressed as a linear combination of these basis elements, $\Gamma\in L^2(I)$.

Any finite subset of the $P_l$ elements will form an orthogonal basis for a finite-dimensional subspace of $L^2(I)$. The HD curve, $\Gamma$, cannot exist completely in one of these finite-dimensional subspaces because it has non-vanishing projections onto an infinite number of the $P_l$ elements. Nonetheless, PTA groups have attempted to reconstruct their interpulsar correlations using a finite number of $P_l$ elements. For instance, see Figure~7 in \cite{aaa+23_a} in which the first six Legendre polynomials are used to reconstruct the ORF. Similar decompositions of the ORF in terms of both Legendre and Chebyshev polynomials can be found in Appendix A of \cite{aaa+23_d}.  Legendre decompositions of the ORF are a justifiable approach to allowing the ORF to be something other than $\Gamma$, important for hypothesis testing and detection of alternative signals. Some truncation in the number of basis elements needs to occur for realistic computational reasons. To assess the scale of error in reconstructing $\Gamma$ induced by this finite truncation, consider that
\begin{eqnarray}
\left(\sum_{l=0}^5 g_l\right)^{-1}\sum_{l=6}^\infty g_l = \frac{3}{32}.
\end{eqnarray}
So all of the HD signal you capture by reconstructing the ORF with just the six lowest-order Legendre polynomials is approximately 10 times larger than the signal that's missed by not including the higher order terms. This is a potentially acceptable concession for computational expediency, particularly since a $\sim$10\% error from an incomplete parameterization of the ORF is currently significantly smaller than the errors on the amplitude of an HD signal just based on current data quality (see the abstract of \cite{aaa+23_a}, for instance). But it is an ultimately unnecessary loss of sensitivity for important science, especially considering the ever-improving nature of PTA data quality---10\% errors are in the near future. We suggest a modified approach. One need not be beholden to Legendre polynomials, even in light of their useful orthogonality properties.  

We will often speak of orthonormal bases as a matter of preference. Normalization is always with respect to a particular inner product. Vectors normalized with respect to the above continuous inner product will be given a ``hat." For example,
\begin{eqnarray}
\hat{\Gamma}(\theta)=\frac{\Gamma(\theta)}{\langle\Gamma,\Gamma\rangle_C^{1/2}}=\sqrt{24}~\Gamma(\theta).
\end{eqnarray}
In later sections, when we use a different discrete inner product in a different vector space, we will adjust our normalization notation accordingly. 

%%%%%%%%%%%%%%%%%%%
%%%%%%%%%%%%%%%%%%%
%%%%%%%%%%%%%%%%%%%
\section{HD-oriented Orthonormal Bases}
Consider the subspace of $L^2(I)$ spanned by the elements of the finite set $S_{\rm in}=\{\Gamma,P_0,P_1,\dots,P_T\}$ for some integer $T$ at which we truncate. This set is a basis for the space it spans as it is linearly independent. However, the presence of $\Gamma$ in the set prevents this basis from being orthogonal so long as $T\geq2$ since $g_l\neq0$ for $l\geq2$, i.e. $\Gamma$ has non-vanishing projections onto other elements of the set if quadruopolar and higher-order Legendre polynomials are included in the set. Fortunately, when given a basis for some space and an inner-product, the Gram-Schmidt (GS) procedure provides a straightforward prescription for generating an orthonormal basis that spans the same space as the original basis. The resultant basis depends on the order in which the original basis elements are fed into the GS algorithm, so whenever we write down a set like $S_{\rm in}$, we assume it to be an ordered set with a first element, a second element, and so forth. We will feed $S_{\rm in}$ into the GS algorithm and generate an orthonormal basis $S_{\rm out}=\{\hat{Q}_0,\hat{Q}_1,\hat{Q}_2,\cdots,\hat{Q}_{T+1}\}$. Because the HD curve is largely quadrupolar ($l=2$) in the sense of Legendre polynomials discussed above, we map $\Gamma$ through this GS procedure to $\hat{Q}_2$. But because the GS algorithm's output depends on the ordering of the input, if we are to preserve HD-correlations in the output set, $\Gamma$ must be the first element of the input set. This leads to some potentially confusing mapping between the elements of $S_{\rm in}$ and $S_{\rm out}$ so to be completely explicit, here's the map of input elements to output elements we employ:

\begin{eqnarray}
\begin{matrix}
\{\Gamma,&P_0,&P_1,&P_2,&\dots,&P_T\}\\
\tikzmarknode{A}{}&\tikzmarknode{B}{}&\tikzmarknode{C}{}&\tikzmarknode{D}{}&&\tikzmarknode{E}{}\\
\tikzmarknode{F}{}&\tikzmarknode{G}{}&\tikzmarknode{H}{}&\tikzmarknode{I}{}&&\tikzmarknode{J}{}\\
\{\hat{Q}_0,&\hat{Q}_1,&\hat{Q}_2,&\hat{Q}_3,&\dots,&\hat{Q}_{T+1},\}.
\end{matrix}
\begin{tikzpicture}[remember picture, overlay]
        \draw[->] (D) -- (I);
        \draw[->] (E) -- (J);
        \draw[->] (A) -- (H);
        \draw[->] (B) -- (F);
        \draw[->] (C) -- (G);
    \end{tikzpicture}
\end{eqnarray}

As a demonstration, we explicitly work through this procedure for the case with $T=3$:

\begin{flalign}
{\rm Step~1a:}~Q_2=\Gamma,\nonumber&&
\end{flalign}
\begin{flalign}
{\rm Step~1b:}~\hat{Q}_2=\frac{Q_2}{\langle Q_2,Q_2\rangle_C^{1/2}}=\boxed{\sqrt{24}\Gamma},&&\\\nonumber
\end{flalign}
\begin{flalign}
{\rm Step~2a:}~Q_0=P_0-\hat{Q}_2\cancelto{0}{\langle P_0,\hat{Q}_2\rangle_C}=P_0,\nonumber&&
\end{flalign}
\begin{flalign}
{\rm Step~2b:}~\hat{Q}_0=\frac{Q_0}{\langle Q_0,Q_0\rangle_C^{1/2}}=\boxed{\frac{1}{\sqrt{2}}P_0},&&\\\nonumber
\end{flalign}
\begin{flalign}
{\rm Step~3a:}~Q_1=P_1-\hat{Q}_2\cancelto{0}{\langle P_1,\hat{Q}_2\rangle_C}-\hat{Q}_0\cancelto{0}{\langle P_1,\hat{Q}_0\rangle_C}=P_1,\nonumber&&
\end{flalign}
\begin{flalign}
{\rm Step~3b:}~\hat{Q}_1=\frac{Q_1}{\langle Q_1,Q_1\rangle_C^{1/2}}=\boxed{\sqrt{\frac{3}{2}}P_1},&&\\\nonumber
\end{flalign}
\begin{flalign}
{\rm Step~4a:}~Q_3=P_2-\hat{Q}_2\langle P_2,\hat{Q}_2\rangle_C-\hat{Q}_1\cancelto{0}{\langle P_2,\hat{Q}_1\rangle_C}\nonumber&&\\-\hat{Q}_0\cancelto{0}{\langle P_2,\hat{Q}_0\rangle_C}=P_2-3\Gamma,\nonumber
\end{flalign}
\begin{flalign}
{\rm Step~4b:}~\hat{Q}_3=\frac{Q_3}{\langle Q_3,Q_3\rangle_C^{1/2}}=\boxed{\sqrt{40}\left(P_2-3\Gamma\right)},&&\\\nonumber
\end{flalign}
\begin{flalign}
{\rm Step~5a:}~Q_4=P_3-\hat{Q}_3\langle P_3,\hat{Q}_3\rangle_C-\hat{Q}_2\langle P_3,\hat{Q}_2\rangle_C\nonumber&&\\-\hat{Q}_1\cancelto{0}{\langle P_3,\hat{Q}_1\rangle_C}-\hat{Q}_0\cancelto{0}{\langle P_3,\hat{Q}_0\rangle_C}\nonumber&&\\=P_3+3P_2-\frac{48}{5}\Gamma,\nonumber
\end{flalign}
\begin{flalign}
{\rm Step~5b:}~\hat{Q}_4=&\frac{Q_4}{\langle Q_4,Q_4\rangle_C^{1/2}}\nonumber\\=&~\boxed{\sqrt{\frac{175}{8}}\left(P_3+3P_2-\frac{48}{5}\Gamma\right)}.
\end{flalign}
Since $\Gamma$ is already orthogonal to $P_0$ and $P_1$ (as evidenced by $g_0=g_1=0$), Steps 1 through 3 of the above procedure are trivial---unit normalization of the input basis functions. Steps 4 and 5 illustrate the general process that emerges for higher-order basis functions. This pattern of steps continues for arbitrarily large choices of $T$. 

The elements of $S_{\rm out}$ are shown in Figure 1 in blue. For the bottom three panels, where $l\geq 2$ and $\hat{Q}_l\neq\hat{P}_l$, we plot the normalized Legendre polynomial (or its negative) with the same number of zero crossings in dashed orange to highlight the differences. The pink curves will be discussed in the next section. In this new basis, HD correlations are purely ``quadrupolar", captured perfectly by $\hat{Q}_2$. By construction, these basis functions are orthonormal with respect to the relevant inner product: $\langle \hat{Q}_k,\hat{Q}_l\rangle_C=\delta_{kl}$. For Legendre polynomials, the number of zero crossings is equal to the order, $l$; this same property carries over to the elements of $S_{\rm out}$, although the locations of the zeros has changed. The even (odd) order Legendre polynomials are even (odd) functions about $\theta=\pi/2$: $P_i\left(\cos{(\theta\pm \pi/2)}\right)=\pm P_i(\cos{\theta})$. It is on this front that $\Gamma$ is most distinct from the $l=2$ Legendre polynomial as it lacks this basic symmetry. For $l\geq 2$, the $\hat{Q}_l$ basis curves also lack this symmetry, reflecting the GW physics at the heart of the HD curve. This is a basis tailored to GW science with PTAs.

%%%%%%%%%%%%%%%%%%%
%%%%%%%%%%%%%%%%%%%
%%%%%%%%%%%%%%%%%%%
\section{Bases Tailored to the Optimal Cross-Correlation Statistic}

Continuous basis functions and the continuous inner product used above are not necessarily the most useful tools possible for PTA data analysis and the reconstruction of ORFs. PTAs sample the ORF on a discrete set of $N=N_p(N_p-1)/2$ angular separations $\theta_i$. For the $i$th pulsar pair, a dimensionless cross-correlation measure $\rho_i$ is measured with an associated uncertainty $\sigma_i$ that depends on the noise properties of the two pulsar timing data sets used to compute it. We will treat the collection of all $N$ cross-correlation measurements and the associated uncertainties as column vectors in an $N$-dimensional vector space, calling them $\rho$ and $\sigma$, respectively. Linear algebra techniques in this vector space are critical to PTA searches for HD correlations. 

\begin{figure}[h!]
    \centering
    \includegraphics[scale=.55]{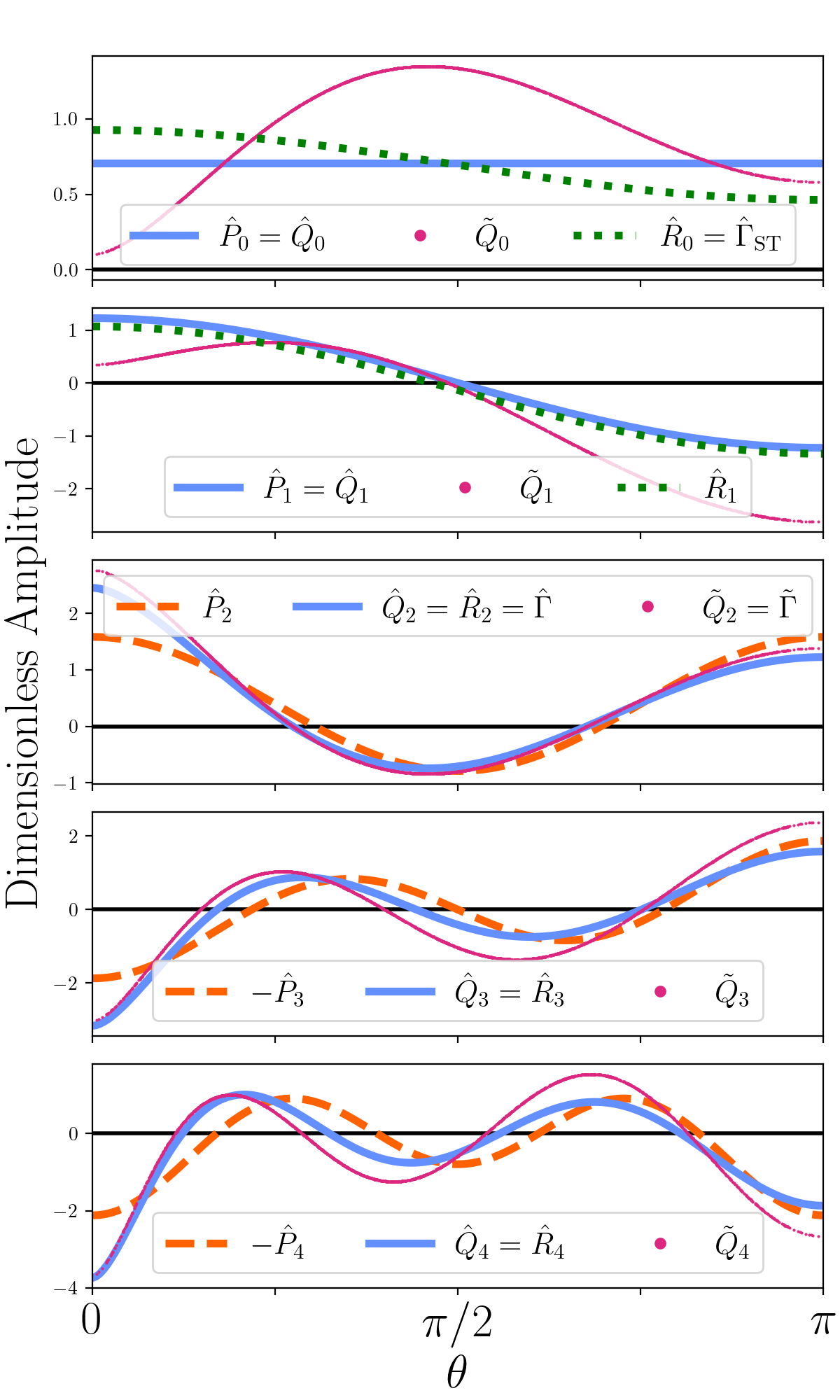}
    \caption{The orthonormal basis elements, $\hat{Q}_l$, built around the normalized Hellings \& Downs (HD) curve, $\hat{Q}_2$, following the procedures described in Section III are displayed in solid blue for $l=0$ through 4. Where these derived basis elements differ from normalized Legendre polynomials ($l=2,~3,$ and $4$), the normalized Legendre polynomial (or its negative) is shown in dashed orange for comparison. The pink ``curves," $\tilde{Q}_l$, are actually discrete collections of $N=2,211$ points, equal to the number of pulsar pairs used for GW searches from NANOGrav's recent 15-year data release \cite{aaa+23_a,aaa+23_e}. These basis elements exist in an $N$-dimensional vector space and were derived through the GS-process using the inner product described in Section~IV. They are designed to search for HD correlations using ``optimal statistic" techniques while allowing for additional structure in the overlap reduction function. Where the $\hat{R}_l$ functions discussed in Section~V differ from the $\hat{Q}_l$ functions, we show them as dotted green curves.}
    \label{fig:idealized}
\end{figure}

Frequentist and quasi-Bayesian searches for HD correlations often make use of the ``optimal statistic" (OS) \cite{dfg+13,ccs+15,vit+18}. Computation of the OS employs estimates of $\rho$ and $\sigma$ that account for the anticipated power-law spectral characteristics of the GW background. Ultimately, the OS is an estimate of the squared amplitude of the background, $A^2_{\rm GW}$, at a particular reference frequency, traditionally $1~{\rm yr}^{-1}$. If we define the diagonal $N\times N$ matrix $\Xi=\sigma \sigma^{\rm T}$, and $\gamma=\Gamma(\theta_i)$, i.e. the $N$-dimensional column vector consisting of the values of the HD curve at the $N$ different angular separations sampled by the PTA, the OS estimate for $A^2_{\rm GW}$ is
\begin{eqnarray}
A^2_{\rm GW}=\frac{\rho^{\rm T}\Xi^{-1}\gamma}{\gamma^{\rm T}\Xi^{-1}\gamma}.
\end{eqnarray}
See Appendix A of \cite{ccs+15} for a discussion of this particular formulation of the OS. 
For our purposes, we recast this formulation of the OS as
\begin{eqnarray}
A^2_{\rm GW} = \frac{\langle\rho,\gamma\rangle_D}{\langle\gamma,\gamma\rangle_D},
\end{eqnarray}
where we have defined the discrete inner product
\begin{eqnarray}
\langle\alpha,\beta\rangle_D=\frac{\alpha^{\rm T}\Xi^{-1}\beta}{{\rm tr}(\Xi^{-1})}.
\end{eqnarray}

Recent work has advanced the OS formalism to accommodate ORFs other than just the HD curve. Specifically, in \cite{sv23}, the ORF is treated as a linear combination of functions $\zeta^i$; these could be Legendre polynomials or something else depending on the chosen application. Rather than a single estimate on the amplitude $A^2_{\rm GW}$ associated with the HD ORF, $\Gamma$, one computes estimates on the multiple squared amplitudes of the multiple correlated processes contributing to the observed correlations:
\begin{eqnarray}
    A^2_i=B_{ij}\langle\rho,\zeta^j\rangle_D,
\end{eqnarray}
where $B_{ij}$ is the matrix inverse to 
\begin{eqnarray}
B^{ij}=\langle\zeta^i,\zeta^j\rangle_D.
\end{eqnarray}
We note that the $A_i^2$ values are sensitive to the scale or amplitude of the basis elements $\zeta^j$ so our preference for unit-normalized basis elements will influence their values.

The GS techniques discussed in the previous section can be advantageously carried over to analyses of this sort. With an input basis $S_{\rm in}=\{\Gamma(\theta_i),P_0(\theta_i),\dots,P_T(\theta_i)\}$ and with the inner product $\langle~,~\rangle_D$, the GS procedure can readily produce an orthonormal basis $S_{\rm out}=\{\tilde{Q}_0,\dots\tilde{Q}_{T+1}\}$, where normalization is now with respect to the discrete inner product.

We have done this and show the results as the pink ``curves" in Fig. 1. They are not continuous since they are only defined on $N$ discrete values of $\theta_i$. We used values of $\theta_i$ and $\sigma_i$ derived from the NANOGrav 15-yr data set \cite{aaa+23_e} where $N_p=67$ and $N=2,211$. Since $N$ is large and the values of $\theta_i$ are fairly well spread through the interval from $0$ to $\pi$, these collections of points look like continuous curves, but some stippling near $\theta=0$ and $\theta=\pi$, populated by fewer pairs of pulsars, betrays the discrete domain. 

If the orthonormal basis we developed here were used in the flexible ORF reconstruction via OS techniques from \cite{sv23}, $B^{ij}=\langle \tilde{Q}_i,\tilde{Q}_j\rangle_D=\delta_{ij}$. The square-amplitude estimates of all of the combined correlated processes become independent of one another with an appropriately constructed basis. Furthermore, $A^2_2$ will be proportional to $A^2_{GW}$ with no spillage of signal power into basis elements of other orders.

Interestingly, one can see from the top two panels of Figure 1 that $\langle\Gamma(\theta_i),P_0(\theta_i)\rangle_D\neq 0$ and $\langle\Gamma(\theta_i),P_1(\theta_i)\rangle_D\neq 0$. In fact, while $\langle \Gamma(\theta_i),P_2(\theta_i)\rangle_D\approx0.089$, $\langle \Gamma(\theta_i),P_0(\theta_i)\rangle_D\approx0.059$ and $\langle \Gamma(\theta_i),P_1(\theta_i)\rangle_D\approx0.051$; the monopole and dipole have substantial overlap with the ORF compared to the anticipated quadrupole. The first few steps of the GS procedure are no longer trivial with this discrete inner product. Thus, $\tilde Q_0\neq\tilde P_0(\theta_i)$ and $Q_1\neq\tilde P_1(\theta_i)$. Purely monopolar and dipolar ORFs are covariant with the HD ORF for NANOGrav and every other PTA. If a PTA group were to look exclusively for HD correlations, the presence of a strong monopolar or dipolar correlation in the data from systematic errors could register as an HD signature; this is among the things directly demonstrated in \cite{thk+16}. The amount of covariance between the HD curve and monopolar and dipolar correlation patterns depends on the specific sampling of $\theta_i$ (which depends on what pulsars a PTA observes) and the uncertainties on cross-correlation measurements (this will vary depending on the specific pulsars, instruments, and observing strategies used by a PTA). In a sense, ${\tilde Q}_0$ and ${\tilde Q}_1$ are the most nearly monopolar and dipolar ORFs, respectively, that NANOGrav could perfectly distinguish from HD correlations without covariances using their 15-yr data set. Also note that $\langle\Gamma,\Gamma\rangle_C\neq\langle\gamma,\gamma\rangle_D$ so $\hat{\Gamma}\neq\tilde{\Gamma}$. This can be seen from the slight mismatch between the blue and pink curves in the middle panel of Figure 1. These two curves are, however, proportional to one another.

%%%%%%%%%%%%%%%%%%%
%%%%%%%%%%%%%%%%%%%
%%%%%%%%%%%%%%%%%%%
\section{Application to Multi-signal Searches}
There are scenarios in which one will want to simultaneously search for specific correlation patterns from multiple different, potentially non-orthogonal processes in a PTA data release. For instance, searches for the HD correlations produced by the two transverse-traceless polarization modes of Einsteinian GWs might be accompanied by a search for evidence of scalar-tensor GW modes with an altogether different ORF \cite{abb+21}. Depending on the inner product being used, these two ORFs may not be orthogonal to one another. A slight modification of the strategies we have been discussing can still facilitate and improve such searches.

If, as an example, we assume the ORF describing the correlations in a particular PTA data release is some linear combination of the HD curve, which we will here call $\Gamma_{\rm HD}$, the ORF appropriate for describing an isotropic background of scalar-tensor GW modes, 
\begin{eqnarray}
\Gamma_{\rm ST}(\mu)=\frac{1}{8}(3+\mu),
\end{eqnarray}
and $T+1$ other linearly-independent basis functions $\eta_i$ (Legendre polynomials, potentially), we can apply the GS procedure with an appropriate inner product to the basis $S_{\rm in}=\{\Gamma_{\rm HD},\Gamma_{\rm ST},\eta_0,\dots\eta_T\}$ to produce an orthonormal basis $S_{\rm out}=\{\bar{R}_0,\dots,\bar{R}_{T+2}\}$ spanning the same space. We are using a bar above basis elements here to represent normalization with respect to whatever inner product is being used. To gain orthonormality, we will potentially have lost a basis function proportional to $\Gamma_{\rm ST}$ in the process (as we will see, $\langle \Gamma_{\rm HD},\Gamma_{ST}\rangle_C = 0$, but $\langle \Gamma_{\rm HD},\Gamma_{ST}\rangle_D \neq 0$ for any realistic PTA data set). This is too great a cost if one were specifically trying to detect or constrain the presence of ST modes. The output of the GS procedure is still valuable for these searches though. 

Suppose that $\bar{R}_2$ and $\bar{R}_0$ are the elements of the GS output basis onto which $\Gamma_{\rm HD}$ and $\Gamma_{\rm ST}$ are mapped, respectively (see following paragraph for why one would do such a mapping). Then the entire part of the original subspace of possible ORFs that is orthogonal to both $\Gamma_{\rm HD}$ and $\Gamma_{\rm ST}$ is spanned by the orthonormal basis $\{\bar{R}_1,\bar{R}_3,\dots\bar{R}_{T+2}\}$. One could simply conduct ORF reconstruction with the basis $\{\bar{\Gamma}_{\rm ST}, \bar{R}_1,\bar{\Gamma}_{\rm HD},\bar{R}_3\dots\bar{R}_{T+2}\}$. This basis is orthonormal but for the fact that $\langle{\Gamma}_{\rm HD},{\Gamma}_{\rm ST}\rangle$, depending on the specific inner product being used, is potentially not equal to zero. Both correlation patterns of interest are completely captured by this basis, the correlations between these two signals can be readily analyzed, and there is a collection of additional orthonormal basis elements for reconstructing any additional structure in the ORF, none of them covariant with the HD and ST signals of interest.

We now explicitly demonstrate this procedure using Legendre polynomials as the additional elements in a five-element basis and the continuous inner product $\langle~,~\rangle_C$. Note that $\Gamma_{\rm ST}$ is a linear combination of $P_0$ and $P_1$. This simplifies much of the needed linear algebra as $\Gamma_{\rm ST}$ is already orthogonal to $\Gamma_{\rm HD}$ and all but two Legendre polynomials. For our input basis, we use $S_{\rm in}=\{\Gamma_{\rm HD},\Gamma_{\rm ST},P_1,P_2,P_3\}$. With $\Gamma_{\rm ST}$ present, including both $P_0$ and $P_1$ in this set would make it linearly dependent, so we exclude $P_0$. In order to preserve the nice feature that Legendre polynomials have whereby the order is equal to the number of zero crossings, we use the map
\begin{eqnarray}
\begin{matrix}
\{\Gamma_{\rm HD},&\Gamma_{ST},&P_1,&P_2,&P_3\}\\
\tikzmarknode{A}{}&\tikzmarknode{B}{}&\tikzmarknode{C}{}&\tikzmarknode{D}{}&\tikzmarknode{E}{}\\
\tikzmarknode{F}{}&\tikzmarknode{G}{}&\tikzmarknode{H}{}&\tikzmarknode{I}{}&\tikzmarknode{J}{}\\
\{\hat{R}_0,&\hat{R}_1,&\hat{R}_2,&\hat{R}_3,&\hat{R}_{4},\}.
\end{matrix}
\begin{tikzpicture}[remember picture, overlay]
        \draw[->] (D) -- (I);
        \draw[->] (E) -- (J);
        \draw[->] (A) -- (H);
        \draw[->] (B) -- (F);
        \draw[->] (C) -- (G);
    \end{tikzpicture}
\end{eqnarray}
The GS procedure with the $\langle~,~\rangle_C$ inner product and the above map produces the orthonormal output basis $S_{\rm out}=\{\hat{R}_0,\hat{R}_1,\hat{R}_2,\hat{R}_3,\hat{R}_4\}$ with
\begin{eqnarray}
\hat{R}_2&=&\hat{Q}_2=\sqrt{24}\Gamma_{\rm HD},\\
\hat{R}_0&=&\sqrt{\frac{24}{7}}\Gamma_{\rm ST},\\
\hat{R}_1&=&\sqrt{\frac{14}{9}}\left(P_1-\frac{2}{7}\Gamma_{\rm ST}\right),\\
\hat{R}_3&=&\hat{Q}_3=\sqrt{40}(P_2-3\Gamma_{\rm HD}),\\
\hat{R}_4&=&\hat{Q}_4=\sqrt{\frac{175}{8}}\left(P_3+3P_2-\frac{48}{5}\Gamma_{\rm HD}\right).
\end{eqnarray}
These $\hat{R}_l$ basis functions differ from the $\hat{Q}_l$ basis functions for only $l=0$ and $l=1$. As such, we explicitly show $\hat{R}_0$ and $\hat{R}_1$ in Figure 1 as dotted green curves. $R_0$ is simply the normalized version of $\Gamma_{\rm ST}$. $\hat{R}_1$ is a slightly rescaled and vertically shifted version of the normalized $l=1$ Legendre polynomial.

If we used this $\hat{R}_l$ basis for ORF reconstruction and if the signal was truly a superposition of just ST and HD signals, all of the power would appear in the $\hat{R}_0$ and $\hat{R}_2$ channels and the interpretation of the result would be very straightforward. If we had used a simple basis of Legendre polynomials for ORF reconstruction, the signal power would be diluted, spread across basis elements of all orders. 

These techniques can be extended to searches for more than two signals present in the data. If one wanted to search for $N_{\rm signal}$ overlapping signals while using $N_{\rm span}$ additional basis elements to span the space of potential additional ORF structure, they would use the set of $N_{\rm signal}+N_{\rm span}$ basis elements as the input set to the orthonormalization procedure (taking care to feed in the $N_{\rm signal}$ elements first and insuring that the input set was linearly independent) and explicitly placing the normalized $N_{\rm signal}$ signal ORFs in the output set. There would potentially be correlations among the signals of interest, but the rest of the output basis would be conveniently orthonormalized around the signal subspace. It is important to adapt one's tools to the task at hand. The techniques described in this paper will prove valuable for flexible and informed ORF reconstruction as PTAs become increasingly powerful over time as diverse probes of the GW Universe. 
%%%%%%%%%%%%%%%%%%%
%%%%%%%%%%%%%%%%%%%
%%%%%%%%%%%%%%%%%%%
\section{Final Remarks}
The techniques we've developed may prove useful for determining the significance of future measurements of HD correlations in yet another way. ``Sky scrambles" are an important and widely-used technique for assessing the false-alarm probability in any search for HD correlations by PTAs \cite{cs16,tlb+17,dzm+23}. The basic idea is to conduct the search for HD correlations using the measured cross correlations and uncertainties, $\rho_i$ and $\sigma_i$, but to artificially rearrange the sky positions of the pulsars so that rather than sampling the correlation space along the true values of pairwise angular separation, $\theta_i$, one samples them along some other set of angular separations, $\theta_i^*$. Not just any sky scramble will do though. One checks that a ``match" statistic, $M$, falls below a sufficiently low threshold to ensure that the scrambled sky is sufficiently different from the true sky to ``kill" the correlated signal. The match statistic used in the literature is a sort of inner product akin to what we have worked with here. If $M\propto\langle\Gamma(\theta_i),\Gamma(\theta_i^*)\rangle$ falls below some tunable threshold, the sky is considered sufficiently scrambled and we look to the value of $\langle\rho_i,\Gamma(\theta_i^*)\rangle$ as being a representative draw from a noise background distribution. 

At its heart, the sky scrambles technique is an effort to assess the distribution of detection statistics against ORFs that are sufficiently orthogonal to the HD ORF. With our techniques, we can generate ORFs that are linear combinations of basis elements which are strictly orthogonal to the HD ORF. We've mapped out the entire space of ORFs orthogonal to HD. We can construct ORFs $\zeta(\theta_i)$ as randomized linear combinations of basis elements that are orthogonal to HD correlations such that $\langle\Gamma(\theta_i),\zeta(\theta_i)\rangle=0$. No scrambling of pulsar positions is necessary to find an ORF that is sufficiently orthogonal to what is expected from a GW signal. Assessing detection statistic distributions in searches for ORFs strictly orthogonal to the HD curve could become an important technique for understanding the noise background of PTAs and making strong detection claims with robust false alarm probabilities in the near future, complementary to sky scrambles and other similar techniques. We leave a detailed exploration of this idea to future work.

\nocite{*}
\bibliography{bespoke}

\begin{thebibliography}{30}%
\makeatletter
\providecommand \@ifxundefined [1]{%
 \@ifx{#1\undefined}
}%
\providecommand \@ifnum [1]{%
 \ifnum #1\expandafter \@firstoftwo
 \else \expandafter \@secondoftwo
 \fi
}%
\providecommand \@ifx [1]{%
 \ifx #1\expandafter \@firstoftwo
 \else \expandafter \@secondoftwo
 \fi
}%
\providecommand \natexlab [1]{#1}%
\providecommand \enquote  [1]{``#1''}%
\providecommand \bibnamefont  [1]{#1}%
\providecommand \bibfnamefont [1]{#1}%
\providecommand \citenamefont [1]{#1}%
\providecommand \href@noop [0]{\@secondoftwo}%
\providecommand \href [0]{\begingroup \@sanitize@url \@href}%
\providecommand \@href[1]{\@@startlink{#1}\@@href}%
\providecommand \@@href[1]{\endgroup#1\@@endlink}%
\providecommand \@sanitize@url [0]{\catcode `\\12\catcode `\$12\catcode
  `\&12\catcode `\#12\catcode `\^12\catcode `\_12\catcode `\%12\relax}%
\providecommand \@@startlink[1]{}%
\providecommand \@@endlink[0]{}%
\providecommand \url  [0]{\begingroup\@sanitize@url \@url }%
\providecommand \@url [1]{\endgroup\@href {#1}{\urlprefix }}%
\providecommand \urlprefix  [0]{URL }%
\providecommand \Eprint [0]{\href }%
\providecommand \doibase [0]{https://doi.org/}%
\providecommand \selectlanguage [0]{\@gobble}%
\providecommand \bibinfo  [0]{\@secondoftwo}%
\providecommand \bibfield  [0]{\@secondoftwo}%
\providecommand \translation [1]{[#1]}%
\providecommand \BibitemOpen [0]{}%
\providecommand \bibitemStop [0]{}%
\providecommand \bibitemNoStop [0]{.\EOS\space}%
\providecommand \EOS [0]{\spacefactor3000\relax}%
\providecommand \BibitemShut  [1]{\csname bibitem#1\endcsname}%
\let\auto@bib@innerbib\@empty
%</preamble>
\bibitem [{\citenamefont {{Ransom}}\ \emph {et~al.}(2019)\citenamefont
  {{Ransom}}, \citenamefont {{Brazier}}, \citenamefont {{Chatterjee}},
  \citenamefont {{Cohen}}, \citenamefont {{Cordes}}, \citenamefont {{DeCesar}},
  \citenamefont {{Demorest}}, \citenamefont {{Hazboun}}, \citenamefont {{Lam}},
  \citenamefont {{Lynch}} \emph {et~al.}}]{rbc+19}%
  \BibitemOpen
  \bibfield  {author} {\bibinfo {author} {\bibfnamefont {S.}~\bibnamefont
  {{Ransom}}}, \bibinfo {author} {\bibfnamefont {A.}~\bibnamefont {{Brazier}}},
  \bibinfo {author} {\bibfnamefont {S.}~\bibnamefont {{Chatterjee}}}, \bibinfo
  {author} {\bibfnamefont {T.}~\bibnamefont {{Cohen}}}, \bibinfo {author}
  {\bibfnamefont {J.~M.}\ \bibnamefont {{Cordes}}}, \bibinfo {author}
  {\bibfnamefont {M.~E.}\ \bibnamefont {{DeCesar}}}, \bibinfo {author}
  {\bibfnamefont {P.~B.}\ \bibnamefont {{Demorest}}}, \bibinfo {author}
  {\bibfnamefont {J.~S.}\ \bibnamefont {{Hazboun}}}, \bibinfo {author}
  {\bibfnamefont {M.~T.}\ \bibnamefont {{Lam}}}, \bibinfo {author}
  {\bibfnamefont {R.~S.}\ \bibnamefont {{Lynch}}}, \emph {et~al.},\ }\bibfield
  {title} {\bibinfo {title} {{The NANOGrav Program for Gravitational Waves and
  Fundamental Physics}},\ }in\ \href
  {https://doi.org/10.48550/arXiv.1908.05356} {\emph {\bibinfo {booktitle}
  {Bulletin of the American Astronomical Society}}},\ Vol.~\bibinfo {volume}
  {51}\ (\bibinfo {year} {2019})\ p.\ \bibinfo {pages} {195},\ \Eprint
  {https://arxiv.org/abs/1908.05356} {arXiv:1908.05356 [astro-ph.IM]}
  \BibitemShut {NoStop}%
\bibitem [{\citenamefont {{Desvignes}}\ \emph {et~al.}(2016)\citenamefont
  {{Desvignes}}, \citenamefont {{Caballero}}, \citenamefont {{Lentati}},
  \citenamefont {{Verbiest}}, \citenamefont {{Champion}}, \citenamefont
  {{Stappers}}, \citenamefont {{Janssen}}, \citenamefont {{Lazarus}},
  \citenamefont {{Os{\l}owski}}, \citenamefont {{Babak}} \emph
  {et~al.}}]{dcl+16}%
  \BibitemOpen
  \bibfield  {author} {\bibinfo {author} {\bibfnamefont {G.}~\bibnamefont
  {{Desvignes}}}, \bibinfo {author} {\bibfnamefont {R.~N.}\ \bibnamefont
  {{Caballero}}}, \bibinfo {author} {\bibfnamefont {L.}~\bibnamefont
  {{Lentati}}}, \bibinfo {author} {\bibfnamefont {J.~P.~W.}\ \bibnamefont
  {{Verbiest}}}, \bibinfo {author} {\bibfnamefont {D.~J.}\ \bibnamefont
  {{Champion}}}, \bibinfo {author} {\bibfnamefont {B.~W.}\ \bibnamefont
  {{Stappers}}}, \bibinfo {author} {\bibfnamefont {G.~H.}\ \bibnamefont
  {{Janssen}}}, \bibinfo {author} {\bibfnamefont {P.}~\bibnamefont
  {{Lazarus}}}, \bibinfo {author} {\bibfnamefont {S.}~\bibnamefont
  {{Os{\l}owski}}}, \bibinfo {author} {\bibfnamefont {S.}~\bibnamefont
  {{Babak}}}, \emph {et~al.},\ }\bibfield  {title} {\bibinfo {title}
  {{High-precision timing of 42 millisecond pulsars with the European Pulsar
  Timing Array}},\ }\href {https://doi.org/10.1093/mnras/stw483} {\bibfield
  {journal} {\bibinfo  {journal} {MNRAS}\ }\textbf {\bibinfo {volume} {458}},\
  \bibinfo {pages} {3341} (\bibinfo {year} {2016})},\ \Eprint
  {https://arxiv.org/abs/1602.08511} {arXiv:1602.08511 [astro-ph.HE]}
  \BibitemShut {NoStop}%
\bibitem [{\citenamefont {{Hobbs}}(2013)}]{h13}%
  \BibitemOpen
  \bibfield  {author} {\bibinfo {author} {\bibfnamefont {G.}~\bibnamefont
  {{Hobbs}}},\ }\bibfield  {title} {\bibinfo {title} {{The Parkes Pulsar Timing
  Array}},\ }\href {https://doi.org/10.1088/0264-9381/30/22/224007} {\bibfield
  {journal} {\bibinfo  {journal} {Classical and Quantum Gravity}\ }\textbf
  {\bibinfo {volume} {30}},\ \bibinfo {eid} {224007} (\bibinfo {year}
  {2013})},\ \Eprint {https://arxiv.org/abs/1307.2629} {arXiv:1307.2629
  [astro-ph.IM]} \BibitemShut {NoStop}%
\bibitem [{\citenamefont {{Lee}}(2016)}]{l16}%
  \BibitemOpen
  \bibfield  {author} {\bibinfo {author} {\bibfnamefont {K.~J.}\ \bibnamefont
  {{Lee}}},\ }\bibfield  {title} {\bibinfo {title} {{Prospects of Gravitational
  Wave Detection Using Pulsar Timing Array for Chinese Future Telescopes}},\
  }in\ \href@noop {} {\emph {\bibinfo {booktitle} {Frontiers in Radio Astronomy
  and FAST Early Sciences Symposium 2015}}},\ \bibinfo {series} {Astronomical
  Society of the Pacific Conference Series}, Vol.\ \bibinfo {volume} {502},\
  \bibinfo {editor} {edited by\ \bibinfo {editor} {\bibfnamefont
  {L.}~\bibnamefont {{Qain}}}\ and\ \bibinfo {editor} {\bibfnamefont
  {D.}~\bibnamefont {{Li}}}}\ (\bibinfo {year} {2016})\ p.~\bibinfo {pages}
  {19}\BibitemShut {NoStop}%
\bibitem [{\citenamefont {{Joshi}}\ \emph {et~al.}(2018)\citenamefont
  {{Joshi}}, \citenamefont {{Arumugasamy}}, \citenamefont {{Bagchi}},
  \citenamefont {{Bandyopadhyay}}, \citenamefont {{Basu}}, \citenamefont
  {{Dhanda Batra}}, \citenamefont {{Bethapudi}}, \citenamefont {{Choudhary}},
  \citenamefont {{De}}, \citenamefont {{Dey}} \emph {et~al.}}]{jab+18}%
  \BibitemOpen
  \bibfield  {author} {\bibinfo {author} {\bibfnamefont {B.~C.}\ \bibnamefont
  {{Joshi}}}, \bibinfo {author} {\bibfnamefont {P.}~\bibnamefont
  {{Arumugasamy}}}, \bibinfo {author} {\bibfnamefont {M.}~\bibnamefont
  {{Bagchi}}}, \bibinfo {author} {\bibfnamefont {D.}~\bibnamefont
  {{Bandyopadhyay}}}, \bibinfo {author} {\bibfnamefont {A.}~\bibnamefont
  {{Basu}}}, \bibinfo {author} {\bibfnamefont {N.}~\bibnamefont {{Dhanda
  Batra}}}, \bibinfo {author} {\bibfnamefont {S.}~\bibnamefont {{Bethapudi}}},
  \bibinfo {author} {\bibfnamefont {A.}~\bibnamefont {{Choudhary}}}, \bibinfo
  {author} {\bibfnamefont {K.}~\bibnamefont {{De}}}, \bibinfo {author}
  {\bibfnamefont {L.}~\bibnamefont {{Dey}}}, \emph {et~al.},\ }\bibfield
  {title} {\bibinfo {title} {{Precision pulsar timing with the ORT and the GMRT
  and its applications in pulsar astrophysics}},\ }\href
  {https://doi.org/10.1007/s12036-018-9549-y} {\bibfield  {journal} {\bibinfo
  {journal} {Journal of Astrophysics and Astronomy}\ }\textbf {\bibinfo
  {volume} {39}},\ \bibinfo {eid} {51} (\bibinfo {year} {2018})}\BibitemShut
  {NoStop}%
\bibitem [{\citenamefont {{Miles}}\ \emph {et~al.}(2023)\citenamefont
  {{Miles}}, \citenamefont {{Shannon}}, \citenamefont {{Bailes}}, \citenamefont
  {{Reardon}}, \citenamefont {{Keith}}, \citenamefont {{Cameron}},
  \citenamefont {{Parthasarathy}}, \citenamefont {{Shamohammadi}},
  \citenamefont {{Spiewak}}, \citenamefont {{van Straten}} \emph
  {et~al.}}]{msb+23}%
  \BibitemOpen
  \bibfield  {author} {\bibinfo {author} {\bibfnamefont {M.~T.}\ \bibnamefont
  {{Miles}}}, \bibinfo {author} {\bibfnamefont {R.~M.}\ \bibnamefont
  {{Shannon}}}, \bibinfo {author} {\bibfnamefont {M.}~\bibnamefont {{Bailes}}},
  \bibinfo {author} {\bibfnamefont {D.~J.}\ \bibnamefont {{Reardon}}}, \bibinfo
  {author} {\bibfnamefont {M.~J.}\ \bibnamefont {{Keith}}}, \bibinfo {author}
  {\bibfnamefont {A.~D.}\ \bibnamefont {{Cameron}}}, \bibinfo {author}
  {\bibfnamefont {A.}~\bibnamefont {{Parthasarathy}}}, \bibinfo {author}
  {\bibfnamefont {M.}~\bibnamefont {{Shamohammadi}}}, \bibinfo {author}
  {\bibfnamefont {R.}~\bibnamefont {{Spiewak}}}, \bibinfo {author}
  {\bibfnamefont {W.}~\bibnamefont {{van Straten}}}, \emph {et~al.},\
  }\bibfield  {title} {\bibinfo {title} {{The MeerKAT Pulsar Timing Array:
  first data release}},\ }\href {https://doi.org/10.1093/mnras/stac3644}
  {\bibfield  {journal} {\bibinfo  {journal} {MNRAS}\ }\textbf {\bibinfo
  {volume} {519}},\ \bibinfo {pages} {3976} (\bibinfo {year} {2023})},\ \Eprint
  {https://arxiv.org/abs/2212.04648} {arXiv:2212.04648 [astro-ph.HE]}
  \BibitemShut {NoStop}%
\bibitem [{\citenamefont {{Verbiest}}\ \emph {et~al.}(2016)\citenamefont
  {{Verbiest}}, \citenamefont {{Lentati}}, \citenamefont {{Hobbs}},
  \citenamefont {{van Haasteren}}, \citenamefont {{Demorest}}, \citenamefont
  {{Janssen}}, \citenamefont {{Wang}}, \citenamefont {{Desvignes}},
  \citenamefont {{Caballero}}, \citenamefont {{Keith}} \emph
  {et~al.}}]{vlh+16}%
  \BibitemOpen
  \bibfield  {author} {\bibinfo {author} {\bibfnamefont {J.~P.~W.}\
  \bibnamefont {{Verbiest}}}, \bibinfo {author} {\bibfnamefont
  {L.}~\bibnamefont {{Lentati}}}, \bibinfo {author} {\bibfnamefont
  {G.}~\bibnamefont {{Hobbs}}}, \bibinfo {author} {\bibfnamefont
  {R.}~\bibnamefont {{van Haasteren}}}, \bibinfo {author} {\bibfnamefont
  {P.~B.}\ \bibnamefont {{Demorest}}}, \bibinfo {author} {\bibfnamefont
  {G.~H.}\ \bibnamefont {{Janssen}}}, \bibinfo {author} {\bibfnamefont {J.~B.}\
  \bibnamefont {{Wang}}}, \bibinfo {author} {\bibfnamefont {G.}~\bibnamefont
  {{Desvignes}}}, \bibinfo {author} {\bibfnamefont {R.~N.}\ \bibnamefont
  {{Caballero}}}, \bibinfo {author} {\bibfnamefont {M.~J.}\ \bibnamefont
  {{Keith}}}, \emph {et~al.},\ }\bibfield  {title} {\bibinfo {title} {{The
  International Pulsar Timing Array: First data release}},\ }\href
  {https://doi.org/10.1093/mnras/stw347} {\bibfield  {journal} {\bibinfo
  {journal} {MNRAS}\ }\textbf {\bibinfo {volume} {458}},\ \bibinfo {pages}
  {1267} (\bibinfo {year} {2016})},\ \Eprint {https://arxiv.org/abs/1602.03640}
  {arXiv:1602.03640 [astro-ph.IM]} \BibitemShut {NoStop}%
\bibitem [{\citenamefont {{Estabrook}}\ and\ \citenamefont
  {{Wahlquist}}(1975)}]{ew75}%
  \BibitemOpen
  \bibfield  {author} {\bibinfo {author} {\bibfnamefont {F.~B.}\ \bibnamefont
  {{Estabrook}}}\ and\ \bibinfo {author} {\bibfnamefont {H.~D.}\ \bibnamefont
  {{Wahlquist}}},\ }\bibfield  {title} {\bibinfo {title} {{Response of Doppler
  spacecraft tracking to gravitational radiation.}},\ }\href
  {https://doi.org/10.1007/BF00762449} {\bibfield  {journal} {\bibinfo
  {journal} {General Relativity and Gravitation}\ }\textbf {\bibinfo {volume}
  {6}},\ \bibinfo {pages} {439} (\bibinfo {year} {1975})}\BibitemShut {NoStop}%
\bibitem [{\citenamefont {{Hellings}}\ and\ \citenamefont
  {{Downs}}(1983)}]{hd83}%
  \BibitemOpen
  \bibfield  {author} {\bibinfo {author} {\bibfnamefont {R.~W.}\ \bibnamefont
  {{Hellings}}}\ and\ \bibinfo {author} {\bibfnamefont {G.~S.}\ \bibnamefont
  {{Downs}}},\ }\bibfield  {title} {\bibinfo {title} {{Upper limits on the
  isotropic gravitational radiation background from pulsar timing analysis.}},\
  }\href {https://doi.org/10.1086/183954} {\bibfield  {journal} {\bibinfo
  {journal} {Astrophysical Journal Letters}\ }\textbf {\bibinfo {volume}
  {265}},\ \bibinfo {pages} {L39} (\bibinfo {year} {1983})}\BibitemShut
  {NoStop}%
\bibitem [{\citenamefont {{Cornish}}\ and\ \citenamefont
  {{Sesana}}(2013)}]{cs13}%
  \BibitemOpen
  \bibfield  {author} {\bibinfo {author} {\bibfnamefont {N.~J.}\ \bibnamefont
  {{Cornish}}}\ and\ \bibinfo {author} {\bibfnamefont {A.}~\bibnamefont
  {{Sesana}}},\ }\bibfield  {title} {\bibinfo {title} {{Pulsar timing array
  analysis for black hole backgrounds}},\ }\href
  {https://doi.org/10.1088/0264-9381/30/22/224005} {\bibfield  {journal}
  {\bibinfo  {journal} {Classical and Quantum Gravity}\ }\textbf {\bibinfo
  {volume} {30}},\ \bibinfo {eid} {224005} (\bibinfo {year}
  {2013})}\BibitemShut {NoStop}%
\bibitem [{\citenamefont {{Allen}}\ and\ \citenamefont
  {{Romano}}(2023)}]{ar22}%
  \BibitemOpen
  \bibfield  {author} {\bibinfo {author} {\bibfnamefont {B.}~\bibnamefont
  {{Allen}}}\ and\ \bibinfo {author} {\bibfnamefont {J.~D.}\ \bibnamefont
  {{Romano}}},\ }\bibfield  {title} {\bibinfo {title} {{Hellings and Downs
  correlation of an arbitrary set of pulsars}},\ }\href
  {https://doi.org/10.1103/PhysRevD.108.043026} {\bibfield  {journal} {\bibinfo
   {journal} {\prd}\ }\textbf {\bibinfo {volume} {108}},\ \bibinfo {eid}
  {043026} (\bibinfo {year} {2023})},\ \Eprint
  {https://arxiv.org/abs/2208.07230} {arXiv:2208.07230 [gr-qc]} \BibitemShut
  {NoStop}%
\bibitem [{\citenamefont {{Reardon}}\ \emph {et~al.}(2023)\citenamefont
  {{Reardon}}, \citenamefont {{Zic}}, \citenamefont {{Shannon}}, \citenamefont
  {{Hobbs}}, \citenamefont {{Bailes}}, \citenamefont {{Di Marco}},
  \citenamefont {{Kapur}}, \citenamefont {{Rogers}}, \citenamefont {{Thrane}},
  \citenamefont {{Askew}} \emph {et~al.}}]{rzs+23}%
  \BibitemOpen
  \bibfield  {author} {\bibinfo {author} {\bibfnamefont {D.~J.}\ \bibnamefont
  {{Reardon}}}, \bibinfo {author} {\bibfnamefont {A.}~\bibnamefont {{Zic}}},
  \bibinfo {author} {\bibfnamefont {R.~M.}\ \bibnamefont {{Shannon}}}, \bibinfo
  {author} {\bibfnamefont {G.~B.}\ \bibnamefont {{Hobbs}}}, \bibinfo {author}
  {\bibfnamefont {M.}~\bibnamefont {{Bailes}}}, \bibinfo {author}
  {\bibfnamefont {V.}~\bibnamefont {{Di Marco}}}, \bibinfo {author}
  {\bibfnamefont {A.}~\bibnamefont {{Kapur}}}, \bibinfo {author} {\bibfnamefont
  {A.~F.}\ \bibnamefont {{Rogers}}}, \bibinfo {author} {\bibfnamefont
  {E.}~\bibnamefont {{Thrane}}}, \bibinfo {author} {\bibfnamefont
  {J.}~\bibnamefont {{Askew}}}, \emph {et~al.},\ }\bibfield  {title} {\bibinfo
  {title} {{Search for an Isotropic Gravitational-wave Background with the
  Parkes Pulsar Timing Array}},\ }\href
  {https://doi.org/10.3847/2041-8213/acdd02} {\bibfield  {journal} {\bibinfo
  {journal} {The Astrophysical Journal Letters}\ }\textbf {\bibinfo {volume}
  {951}},\ \bibinfo {eid} {L6} (\bibinfo {year} {2023})},\ \Eprint
  {https://arxiv.org/abs/2306.16215} {arXiv:2306.16215 [astro-ph.HE]}
  \BibitemShut {NoStop}%
\bibitem [{\citenamefont {{Xu}}\ \emph {et~al.}(2023)\citenamefont {{Xu}},
  \citenamefont {{Chen}}, \citenamefont {{Guo}}, \citenamefont {{Jiang}},
  \citenamefont {{Wang}}, \citenamefont {{Xu}}, \citenamefont {{Xue}},
  \citenamefont {{Nicolas Caballero}}, \citenamefont {{Yuan}}, \citenamefont
  {{Xu}} \emph {et~al.}}]{xcg+23}%
  \BibitemOpen
  \bibfield  {author} {\bibinfo {author} {\bibfnamefont {H.}~\bibnamefont
  {{Xu}}}, \bibinfo {author} {\bibfnamefont {S.}~\bibnamefont {{Chen}}},
  \bibinfo {author} {\bibfnamefont {Y.}~\bibnamefont {{Guo}}}, \bibinfo
  {author} {\bibfnamefont {J.}~\bibnamefont {{Jiang}}}, \bibinfo {author}
  {\bibfnamefont {B.}~\bibnamefont {{Wang}}}, \bibinfo {author} {\bibfnamefont
  {J.}~\bibnamefont {{Xu}}}, \bibinfo {author} {\bibfnamefont {Z.}~\bibnamefont
  {{Xue}}}, \bibinfo {author} {\bibfnamefont {R.}~\bibnamefont {{Nicolas
  Caballero}}}, \bibinfo {author} {\bibfnamefont {J.}~\bibnamefont {{Yuan}}},
  \bibinfo {author} {\bibfnamefont {Y.}~\bibnamefont {{Xu}}}, \emph {et~al.},\
  }\bibfield  {title} {\bibinfo {title} {{Searching for the Nano-Hertz
  Stochastic Gravitational Wave Background with the Chinese Pulsar Timing Array
  Data Release I}},\ }\href {https://doi.org/10.1088/1674-4527/acdfa5}
  {\bibfield  {journal} {\bibinfo  {journal} {Research in Astronomy and
  Astrophysics}\ }\textbf {\bibinfo {volume} {23}},\ \bibinfo {eid} {075024}
  (\bibinfo {year} {2023})},\ \Eprint {https://arxiv.org/abs/2306.16216}
  {arXiv:2306.16216 [astro-ph.HE]} \BibitemShut {NoStop}%
\bibitem [{\citenamefont {{Antoniadis \emph{et al.}}}\ \emph
  {et~al.}(2023)\citenamefont {{Antoniadis \emph{et al.}}}, \citenamefont
  {{EPTA Collaboration}},\ and\ \citenamefont {{InPTA
  Collaboration}}}]{aaa+23_d}%
  \BibitemOpen
  \bibfield  {author} {\bibinfo {author} {\bibfnamefont {J.}~\bibnamefont
  {{Antoniadis \emph{et al.}}}}, \bibinfo {author} {\bibnamefont {{EPTA
  Collaboration}}},\ and\ \bibinfo {author} {\bibnamefont {{InPTA
  Collaboration}}},\ }\bibfield  {title} {\bibinfo {title} {{The second data
  release from the European Pulsar Timing Array. III. Search for gravitational
  wave signals}},\ }\href {https://doi.org/10.1051/0004-6361/202346844}
  {\bibfield  {journal} {\bibinfo  {journal} {Astronomy \& Astrophysics}\
  }\textbf {\bibinfo {volume} {678}},\ \bibinfo {eid} {A50} (\bibinfo {year}
  {2023})},\ \Eprint {https://arxiv.org/abs/2306.16214} {arXiv:2306.16214
  [astro-ph.HE]} \BibitemShut {NoStop}%
\bibitem [{\citenamefont {{Agazie \emph{et al.},}}\ and\ \citenamefont
  {{NANOGrav Collaboration}}(2023{\natexlab{a}})}]{aaa+23_a}%
  \BibitemOpen
  \bibfield  {author} {\bibinfo {author} {\bibfnamefont {G.}~\bibnamefont
  {{Agazie \emph{et al.},}}}\ and\ \bibinfo {author} {\bibnamefont {{NANOGrav
  Collaboration}}},\ }\bibfield  {title} {\bibinfo {title} {{The NANOGrav 15 yr
  Data Set: Evidence for a Gravitational-wave Background}},\ }\href
  {https://doi.org/10.3847/2041-8213/acdac6} {\bibfield  {journal} {\bibinfo
  {journal} {The Astrophysical Journal Letters}\ }\textbf {\bibinfo {volume}
  {951}},\ \bibinfo {eid} {L8} (\bibinfo {year}
  {2023}{\natexlab{a}})}\BibitemShut {NoStop}%
\bibitem [{\citenamefont {{Agazie \emph{et al.},}}\ and\ \citenamefont
  {{NANOGrav Collaboration}}(2023{\natexlab{b}})}]{aaa+23_b}%
  \BibitemOpen
  \bibfield  {author} {\bibinfo {author} {\bibfnamefont {G.}~\bibnamefont
  {{Agazie \emph{et al.},}}}\ and\ \bibinfo {author} {\bibnamefont {{NANOGrav
  Collaboration}}},\ }\bibfield  {title} {\bibinfo {title} {{The NANOGrav 15 yr
  Data Set: Constraints on Supermassive Black Hole Binaries from the
  Gravitational-wave Background}},\ }\href
  {https://doi.org/10.3847/2041-8213/ace18b} {\bibfield  {journal} {\bibinfo
  {journal} {The Astrophysical Journal Letters}\ }\textbf {\bibinfo {volume}
  {952}},\ \bibinfo {eid} {L37} (\bibinfo {year} {2023}{\natexlab{b}})},\
  \Eprint {https://arxiv.org/abs/2306.16220} {arXiv:2306.16220 [astro-ph.HE]}
  \BibitemShut {NoStop}%
\bibitem [{\citenamefont {{Afzal \emph{et al.},}}\ and\ \citenamefont
  {{NANOGrav Collaboration}}(2023)}]{aaa+23_c}%
  \BibitemOpen
  \bibfield  {author} {\bibinfo {author} {\bibfnamefont {A.}~\bibnamefont
  {{Afzal \emph{et al.},}}}\ and\ \bibinfo {author} {\bibnamefont {{NANOGrav
  Collaboration}}},\ }\bibfield  {title} {\bibinfo {title} {{The NANOGrav 15 yr
  Data Set: Search for Signals from New Physics}},\ }\href
  {https://doi.org/10.3847/2041-8213/acdc91} {\bibfield  {journal} {\bibinfo
  {journal} {The Astrophysical Journal Letters}\ }\textbf {\bibinfo {volume}
  {951}},\ \bibinfo {eid} {L11} (\bibinfo {year} {2023})}\BibitemShut {NoStop}%
\bibitem [{\citenamefont {{Tiburzi}}\ \emph {et~al.}(2016)\citenamefont
  {{Tiburzi}}, \citenamefont {{Hobbs}}, \citenamefont {{Kerr}}, \citenamefont
  {{Coles}}, \citenamefont {{Dai}}, \citenamefont {{Manchester}}, \citenamefont
  {{Possenti}}, \citenamefont {{Shannon}},\ and\ \citenamefont
  {{You}}}]{thk+16}%
  \BibitemOpen
  \bibfield  {author} {\bibinfo {author} {\bibfnamefont {C.}~\bibnamefont
  {{Tiburzi}}}, \bibinfo {author} {\bibfnamefont {G.}~\bibnamefont {{Hobbs}}},
  \bibinfo {author} {\bibfnamefont {M.}~\bibnamefont {{Kerr}}}, \bibinfo
  {author} {\bibfnamefont {W.~A.}\ \bibnamefont {{Coles}}}, \bibinfo {author}
  {\bibfnamefont {S.}~\bibnamefont {{Dai}}}, \bibinfo {author} {\bibfnamefont
  {R.~N.}\ \bibnamefont {{Manchester}}}, \bibinfo {author} {\bibfnamefont
  {A.}~\bibnamefont {{Possenti}}}, \bibinfo {author} {\bibfnamefont {R.~M.}\
  \bibnamefont {{Shannon}}},\ and\ \bibinfo {author} {\bibfnamefont {X.~P.}\
  \bibnamefont {{You}}},\ }\bibfield  {title} {\bibinfo {title} {{A study of
  spatial correlations in pulsar timing array data}},\ }\href
  {https://doi.org/10.1093/mnras/stv2143} {\bibfield  {journal} {\bibinfo
  {journal} {MNRAS}\ }\textbf {\bibinfo {volume} {455}},\ \bibinfo {pages}
  {4339} (\bibinfo {year} {2016})}\BibitemShut {NoStop}%
\bibitem [{\citenamefont {{Chamberlin}}\ and\ \citenamefont
  {{Siemens}}(2012)}]{cs12}%
  \BibitemOpen
  \bibfield  {author} {\bibinfo {author} {\bibfnamefont {S.~J.}\ \bibnamefont
  {{Chamberlin}}}\ and\ \bibinfo {author} {\bibfnamefont {X.}~\bibnamefont
  {{Siemens}}},\ }\bibfield  {title} {\bibinfo {title} {{Stochastic backgrounds
  in alternative theories of gravity: Overlap reduction functions for pulsar
  timing arrays}},\ }\href {https://doi.org/10.1103/PhysRevD.85.082001}
  {\bibfield  {journal} {\bibinfo  {journal} {\prd}\ }\textbf {\bibinfo
  {volume} {85}},\ \bibinfo {eid} {082001} (\bibinfo {year}
  {2012})}\BibitemShut {NoStop}%
\bibitem [{\citenamefont {{Gair}}\ \emph {et~al.}(2015)\citenamefont {{Gair}},
  \citenamefont {{Romano}},\ and\ \citenamefont {{Taylor}}}]{grt15}%
  \BibitemOpen
  \bibfield  {author} {\bibinfo {author} {\bibfnamefont {J.~R.}\ \bibnamefont
  {{Gair}}}, \bibinfo {author} {\bibfnamefont {J.~D.}\ \bibnamefont
  {{Romano}}},\ and\ \bibinfo {author} {\bibfnamefont {S.~R.}\ \bibnamefont
  {{Taylor}}},\ }\bibfield  {title} {\bibinfo {title} {{Mapping
  gravitational-wave backgrounds of arbitrary polarisation using pulsar timing
  arrays}},\ }\href {https://doi.org/10.1103/PhysRevD.92.102003} {\bibfield
  {journal} {\bibinfo  {journal} {\prd}\ }\textbf {\bibinfo {volume} {92}},\
  \bibinfo {eid} {102003} (\bibinfo {year} {2015})}\BibitemShut {NoStop}%
\bibitem [{\citenamefont {{Gair}}\ \emph {et~al.}(2014)\citenamefont {{Gair}},
  \citenamefont {{Romano}}, \citenamefont {{Taylor}},\ and\ \citenamefont
  {{Mingarelli}}}]{grt+14}%
  \BibitemOpen
  \bibfield  {author} {\bibinfo {author} {\bibfnamefont {J.}~\bibnamefont
  {{Gair}}}, \bibinfo {author} {\bibfnamefont {J.~D.}\ \bibnamefont
  {{Romano}}}, \bibinfo {author} {\bibfnamefont {S.}~\bibnamefont {{Taylor}}},\
  and\ \bibinfo {author} {\bibfnamefont {C.~M.~F.}\ \bibnamefont
  {{Mingarelli}}},\ }\bibfield  {title} {\bibinfo {title} {{Mapping
  gravitational-wave backgrounds using methods from CMB analysis: Application
  to pulsar timing arrays}},\ }\href
  {https://doi.org/10.1103/PhysRevD.90.082001} {\bibfield  {journal} {\bibinfo
  {journal} {\prd}\ }\textbf {\bibinfo {volume} {90}},\ \bibinfo {eid} {082001}
  (\bibinfo {year} {2014})}\BibitemShut {NoStop}%
\bibitem [{\citenamefont {{Agazie \emph{et al.},}}\ and\ \citenamefont
  {{NANOGrav Collaboration}}(2023{\natexlab{c}})}]{aaa+23_e}%
  \BibitemOpen
  \bibfield  {author} {\bibinfo {author} {\bibfnamefont {G.}~\bibnamefont
  {{Agazie \emph{et al.},}}}\ and\ \bibinfo {author} {\bibnamefont {{NANOGrav
  Collaboration}}},\ }\bibfield  {title} {\bibinfo {title} {{The NANOGrav 15 yr
  Data Set: Observations and Timing of 68 Millisecond Pulsars}},\ }\href
  {https://doi.org/10.3847/2041-8213/acda9a} {\bibfield  {journal} {\bibinfo
  {journal} {The Astrophysical Journal Letters}\ }\textbf {\bibinfo {volume}
  {951}},\ \bibinfo {eid} {L9} (\bibinfo {year}
  {2023}{\natexlab{c}})}\BibitemShut {NoStop}%
\bibitem [{\citenamefont {{Demorest}}\ \emph {et~al.}(2013)\citenamefont
  {{Demorest}}, \citenamefont {{Ferdman}}, \citenamefont {{Gonzalez}},
  \citenamefont {{Nice}}, \citenamefont {{Ransom}}, \citenamefont {{Stairs}},
  \citenamefont {{Arzoumanian}}, \citenamefont {{Brazier}}, \citenamefont
  {{Burke-Spolaor}}, \citenamefont {{Chamberlin}} \emph {et~al.}}]{dfg+13}%
  \BibitemOpen
  \bibfield  {author} {\bibinfo {author} {\bibfnamefont {P.~B.}\ \bibnamefont
  {{Demorest}}}, \bibinfo {author} {\bibfnamefont {R.~D.}\ \bibnamefont
  {{Ferdman}}}, \bibinfo {author} {\bibfnamefont {M.~E.}\ \bibnamefont
  {{Gonzalez}}}, \bibinfo {author} {\bibfnamefont {D.}~\bibnamefont {{Nice}}},
  \bibinfo {author} {\bibfnamefont {S.}~\bibnamefont {{Ransom}}}, \bibinfo
  {author} {\bibfnamefont {I.~H.}\ \bibnamefont {{Stairs}}}, \bibinfo {author}
  {\bibfnamefont {Z.}~\bibnamefont {{Arzoumanian}}}, \bibinfo {author}
  {\bibfnamefont {A.}~\bibnamefont {{Brazier}}}, \bibinfo {author}
  {\bibfnamefont {S.}~\bibnamefont {{Burke-Spolaor}}}, \bibinfo {author}
  {\bibfnamefont {S.~J.}\ \bibnamefont {{Chamberlin}}}, \emph {et~al.},\
  }\bibfield  {title} {\bibinfo {title} {{Limits on the Stochastic
  Gravitational Wave Background from the North American Nanohertz Observatory
  for Gravitational Waves}},\ }\href
  {https://doi.org/10.1088/0004-637X/762/2/94} {\bibfield  {journal} {\bibinfo
  {journal} {\apj}\ }\textbf {\bibinfo {volume} {762}},\ \bibinfo {eid} {94}
  (\bibinfo {year} {2013})},\ \Eprint {https://arxiv.org/abs/1201.6641}
  {arXiv:1201.6641 [astro-ph.CO]} \BibitemShut {NoStop}%
\bibitem [{\citenamefont {{Chamberlin}}\ \emph {et~al.}(2015)\citenamefont
  {{Chamberlin}}, \citenamefont {{Creighton}}, \citenamefont {{Siemens}},
  \citenamefont {{Demorest}}, \citenamefont {{Ellis}}, \citenamefont
  {{Price}},\ and\ \citenamefont {{Romano}}}]{ccs+15}%
  \BibitemOpen
  \bibfield  {author} {\bibinfo {author} {\bibfnamefont {S.~J.}\ \bibnamefont
  {{Chamberlin}}}, \bibinfo {author} {\bibfnamefont {J.~D.~E.}\ \bibnamefont
  {{Creighton}}}, \bibinfo {author} {\bibfnamefont {X.}~\bibnamefont
  {{Siemens}}}, \bibinfo {author} {\bibfnamefont {P.}~\bibnamefont
  {{Demorest}}}, \bibinfo {author} {\bibfnamefont {J.}~\bibnamefont {{Ellis}}},
  \bibinfo {author} {\bibfnamefont {L.~R.}\ \bibnamefont {{Price}}},\ and\
  \bibinfo {author} {\bibfnamefont {J.~D.}\ \bibnamefont {{Romano}}},\
  }\bibfield  {title} {\bibinfo {title} {{Time-domain implementation of the
  optimal cross-correlation statistic for stochastic gravitational-wave
  background searches in pulsar timing data}},\ }\href
  {https://doi.org/10.1103/PhysRevD.91.044048} {\bibfield  {journal} {\bibinfo
  {journal} {\prd}\ }\textbf {\bibinfo {volume} {91}},\ \bibinfo {eid} {044048}
  (\bibinfo {year} {2015})},\ \Eprint {https://arxiv.org/abs/1410.8256}
  {arXiv:1410.8256 [astro-ph.IM]} \BibitemShut {NoStop}%
\bibitem [{\citenamefont {{Vigeland}}\ \emph {et~al.}(2018)\citenamefont
  {{Vigeland}}, \citenamefont {{Islo}}, \citenamefont {{Taylor}},\ and\
  \citenamefont {{Ellis}}}]{vit+18}%
  \BibitemOpen
  \bibfield  {author} {\bibinfo {author} {\bibfnamefont {S.~J.}\ \bibnamefont
  {{Vigeland}}}, \bibinfo {author} {\bibfnamefont {K.}~\bibnamefont {{Islo}}},
  \bibinfo {author} {\bibfnamefont {S.~R.}\ \bibnamefont {{Taylor}}},\ and\
  \bibinfo {author} {\bibfnamefont {J.~A.}\ \bibnamefont {{Ellis}}},\
  }\bibfield  {title} {\bibinfo {title} {{Noise-marginalized optimal statistic:
  A robust hybrid frequentist-Bayesian statistic for the stochastic
  gravitational-wave background in pulsar timing arrays}},\ }\href
  {https://doi.org/10.1103/PhysRevD.98.044003} {\bibfield  {journal} {\bibinfo
  {journal} {\prd}\ }\textbf {\bibinfo {volume} {98}},\ \bibinfo {eid} {044003}
  (\bibinfo {year} {2018})}\BibitemShut {NoStop}%
\bibitem [{\citenamefont {{Sardesai}}\ \emph {et~al.}(2023)\citenamefont
  {{Sardesai}}, \citenamefont {{Vigeland}}, \citenamefont {{Gersbach}},\ and\
  \citenamefont {{Taylor}}}]{sv23}%
  \BibitemOpen
  \bibfield  {author} {\bibinfo {author} {\bibfnamefont {S.~C.}\ \bibnamefont
  {{Sardesai}}}, \bibinfo {author} {\bibfnamefont {S.~J.}\ \bibnamefont
  {{Vigeland}}}, \bibinfo {author} {\bibfnamefont {K.~A.}\ \bibnamefont
  {{Gersbach}}},\ and\ \bibinfo {author} {\bibfnamefont {S.~R.}\ \bibnamefont
  {{Taylor}}},\ }\bibfield  {title} {\bibinfo {title} {{Generalized optimal
  statistic for characterizing multiple correlated signals in pulsar timing
  arrays}},\ }\href {https://doi.org/10.1103/PhysRevD.108.124081} {\bibfield
  {journal} {\bibinfo  {journal} {\prd}\ }\textbf {\bibinfo {volume} {108}},\
  \bibinfo {eid} {124081} (\bibinfo {year} {2023})},\ \Eprint
  {https://arxiv.org/abs/2303.09615} {arXiv:2303.09615 [astro-ph.IM]}
  \BibitemShut {NoStop}%
\bibitem [{\citenamefont {{Arzoumanian \emph{et al.},}}\ and\ \citenamefont
  {{NANOGrav Collaboration}}(2021)}]{abb+21}%
  \BibitemOpen
  \bibfield  {author} {\bibinfo {author} {\bibfnamefont {Z.}~\bibnamefont
  {{Arzoumanian \emph{et al.},}}}\ and\ \bibinfo {author} {\bibnamefont
  {{NANOGrav Collaboration}}},\ }\bibfield  {title} {\bibinfo {title} {{The
  NANOGrav 12.5-year Data Set: Search for Non-Einsteinian Polarization Modes in
  the Gravitational-wave Background}},\ }\href
  {https://doi.org/10.3847/2041-8213/ac401c} {\bibfield  {journal} {\bibinfo
  {journal} {The Astrophysical Journal Letters}\ }\textbf {\bibinfo {volume}
  {923}},\ \bibinfo {eid} {L22} (\bibinfo {year} {2021})}\BibitemShut {NoStop}%
\bibitem [{\citenamefont {{Cornish}}\ and\ \citenamefont
  {{Sampson}}(2016)}]{cs16}%
  \BibitemOpen
  \bibfield  {author} {\bibinfo {author} {\bibfnamefont {N.~J.}\ \bibnamefont
  {{Cornish}}}\ and\ \bibinfo {author} {\bibfnamefont {L.}~\bibnamefont
  {{Sampson}}},\ }\bibfield  {title} {\bibinfo {title} {{Towards robust
  gravitational wave detection with pulsar timing arrays}},\ }\href
  {https://doi.org/10.1103/PhysRevD.93.104047} {\bibfield  {journal} {\bibinfo
  {journal} {\prd}\ }\textbf {\bibinfo {volume} {93}},\ \bibinfo {eid} {104047}
  (\bibinfo {year} {2016})}\BibitemShut {NoStop}%
\bibitem [{\citenamefont {{Taylor}}\ \emph {et~al.}(2017)\citenamefont
  {{Taylor}}, \citenamefont {{Lentati}}, \citenamefont {{Babak}}, \citenamefont
  {{Brem}}, \citenamefont {{Gair}}, \citenamefont {{Sesana}},\ and\
  \citenamefont {{Vecchio}}}]{tlb+17}%
  \BibitemOpen
  \bibfield  {author} {\bibinfo {author} {\bibfnamefont {S.~R.}\ \bibnamefont
  {{Taylor}}}, \bibinfo {author} {\bibfnamefont {L.}~\bibnamefont {{Lentati}}},
  \bibinfo {author} {\bibfnamefont {S.}~\bibnamefont {{Babak}}}, \bibinfo
  {author} {\bibfnamefont {P.}~\bibnamefont {{Brem}}}, \bibinfo {author}
  {\bibfnamefont {J.~R.}\ \bibnamefont {{Gair}}}, \bibinfo {author}
  {\bibfnamefont {A.}~\bibnamefont {{Sesana}}},\ and\ \bibinfo {author}
  {\bibfnamefont {A.}~\bibnamefont {{Vecchio}}},\ }\bibfield  {title} {\bibinfo
  {title} {{All correlations must die: Assessing the significance of a
  stochastic gravitational-wave background in pulsar timing arrays}},\ }\href
  {https://doi.org/10.1103/PhysRevD.95.042002} {\bibfield  {journal} {\bibinfo
  {journal} {\prd}\ }\textbf {\bibinfo {volume} {95}},\ \bibinfo {eid} {042002}
  (\bibinfo {year} {2017})}\BibitemShut {NoStop}%
\bibitem [{\citenamefont {{Di Marco}}\ \emph {et~al.}(2023)\citenamefont {{Di
  Marco}}, \citenamefont {{Zic}}, \citenamefont {{Miles}}, \citenamefont
  {{Reardon}}, \citenamefont {{Thrane}},\ and\ \citenamefont
  {{Shannon}}}]{dzm+23}%
  \BibitemOpen
  \bibfield  {author} {\bibinfo {author} {\bibfnamefont {V.}~\bibnamefont {{Di
  Marco}}}, \bibinfo {author} {\bibfnamefont {A.}~\bibnamefont {{Zic}}},
  \bibinfo {author} {\bibfnamefont {M.~T.}\ \bibnamefont {{Miles}}}, \bibinfo
  {author} {\bibfnamefont {D.~J.}\ \bibnamefont {{Reardon}}}, \bibinfo {author}
  {\bibfnamefont {E.}~\bibnamefont {{Thrane}}},\ and\ \bibinfo {author}
  {\bibfnamefont {R.~M.}\ \bibnamefont {{Shannon}}},\ }\bibfield  {title}
  {\bibinfo {title} {{Toward Robust Detections of Nanohertz Gravitational
  Waves}},\ }\href {https://doi.org/10.3847/1538-4357/acee71} {\bibfield
  {journal} {\bibinfo  {journal} {\apj}\ }\textbf {\bibinfo {volume} {956}},\
  \bibinfo {eid} {14} (\bibinfo {year} {2023})},\ \Eprint
  {https://arxiv.org/abs/2305.04464} {arXiv:2305.04464 [astro-ph.IM]}
  \BibitemShut {NoStop}%
\end{thebibliography}%
\end{document}